# Transportation and Systems Analysis Collaborations in Support of a Federal Consolidated Interim Storage Facility

Robby Joseph[1], Harish Gadey[2], Brian Craig[3], Lucas Vander Wal[3], Mark Abkowitz[4], Robert Claypool[5],
Riley Cumberland[6], Caitlin Condon[2], Miriam Juckett[2], Steve Maheras[2], Gordon Petersen[1]

[1]Idaho National Laboratory, Idaho Falls, ID
[2]Pacific Northwest National Laboratory, Richland, WA
[3]Argonne National Laboratory, Lemont, IL
[4]Vanderbilt University, Nashville, TN
[5]Kanini, Nashville, TN
[6]Oak Ridge National Laboratory, Oak Ridge, TN

**ABSTRACT**

The U.S. Department of Energy's Integrated Waste Management (IWM) program under the Office of Nuclear Energy is planning for the future transportation, storage, and eventual disposal of spent nuclear fuel (SNF) and high-level radioactive waste (HLW) from nuclear power plant and waste custodian sites across the United States. To better enable informed decision-making regarding the back end of the nuclear fuel cycle, the IWM program has been sponsoring the development and application of system analysis tools capable of analyzing various options for managing SNF and HLW. With these tools, integrated waste management system (IWMS) architecture analyses are being conducted to support the future deployment of a comprehensive nuclear waste management system that considers all major back-end aspects of the nuclear fuel cycle (i.e., transportation, storage, and disposal). System analyses and assessments typically use these modeling and simulation tools to investigate implications of changes in various assumptions and parameters such as acceptance rates, receipt logic, facility capacities and capabilities, use of standardized canisters, start and stop dates of facilities, etc.

The Next Generation System Analysis Model (NGSAM) is an agent-based simulation toolkit that is used for a system-level simulation and analysis focused on SNF management in the United States. NGSAM's primary purpose is to provide a system analyst with capabilities to model the IWMS and gain insights into SNF and HLW management alternatives including the impact of system choices, associated cost estimates, and development of integrated yet flexible approaches. An analyst using NGSAM can define several factors like the number of storage facilities, capacity at each facility, transportation schedules, shipment rates, and other conditions.

IWM is also developing the Stakeholder Tool for Assessing Radioactive Transportation (START). START is a web-based geospatial tool developed to provide the visualization and initial evaluation of transportation options associated with future SNF and HLW shipment planning and operations. This includes characterizing safety, economic, and environmental conditions at and in proximity to shipment origins as well as along prospective transportation routes. Information from the START tool can be presented/shared as maps, graphics, geospatial files, and tabular form to enable easy data representation and export functionality.

The system analysis, NGSAM, and START teams have been working closely for several years before formalizing this collaboration. START provides the SNF transportation routing information for use in NGSAM. System analysts use the NGSAM tool to generate results (schedule, costs, infrastructure acquisitions, shipment rates, etc.). Subject-matter experts and system analysts work together to inform how NGSAM should model the waste management system. Specifically, this paper discusses the collaborations between the system analysis, NGSAM, and START teams and their accomplishments over the past year. Some examples of these efforts include calibrating and adding data to the START output files to meet NGSAM needs, gaining a better understanding of START data used in NGSAM as well as how updates in START data could result in an improved NGSAM analysis. Detailed examples of various tasks that have been performed by the team will be discussed in the full paper. Continued efforts in this direction are expected in the coming years.

**INTRODUCTION**

The U.S. Department of Energy (DOE) Integrated Waste Management (IWM) program under the Office of Nuclear Energy is planning for the future transportation, storage, and eventual disposal of spent nuclear fuel (SNF) and high-level radioactive waste (HLW) from nuclear power plant and waste custodian sites across the United States. To better enable informed decision-making regarding the back end of the nuclear fuel cycle, the IWM program has been sponsoring the development and application of system analysis tools capable of analyzing various options for managing SNF and HLW. With these tools, integrated waste management system (IWMS) architecture analyses are being conducted to support the future deployment of a comprehensive nuclear waste management system that considers all major back-end aspects of the nuclear fuel cycle (i.e., transportation, storage, and disposal). System analyses and assessments typically use these modeling and simulation tools to investigate implications of changes in various assumptions and parameters such as acceptance rates, receipt logic, facility capacities and capabilities, use of standardized canisters, start and stop dates of facilities, etc[1].

**Systems Analysis**

At the end of 2023, it is expected that there will be 92 operating and 27 shutdown commercial light water reactors in the United States with SNF stored in either wet (pool) or dry (canister-overpack systems) storage [1, 2]. Moving the U.S. SNF inventory from multiple locations to either a consolidated interim storage facility and/or a permanent disposal facility requires applying systems-level engineering. The IWM program is applying waste management architecture analysis and decision analysis principles to inform the design of the IWMS, which is a system of systems. The desired outcome is to gain knowledge and develop an integrated analysis approach that will be implemented on the three major systems of the fuel-cycle's back end, namely transportation, storage, and disposal, to design a successful IWMS (Figure 1).

---

[1] Disclaimer: This is a technical paper that does not take into account contractual limitations or obligations under the Standard Contract for Disposal of Spent Nuclear Fuel and/or High-Level Radioactive Waste (Standard Contract) (10 CFR Part 961). To the extent discussions or recommendations in this paper conflict with the provisions of the Standard Contract, the Standard Contract governs the obligations of the parties, and this paper in no manner supersedes, overrides, or amends the Standard Contract. This paper reflects technical work which could support future decision making by DOE. No inferences should be drawn from this paper regarding future actions by DOE, which are limited both by the terms of the Standard Contract and Congressional appropriations for the Department to fulfill its obligations under the Nuclear Waste Policy Act, including licensing and construction of a spent nuclear fuel repository.

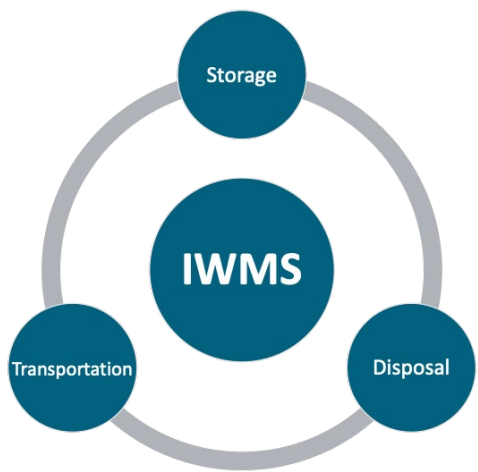

Figure. 1. The three major systems integral to an over-arching IWMS.

Transportation of SNF has occurred by truck, barge, and rail. Commercial SNF transportation packages weigh between 80 and 210 tons [3], exceeding the legal weight limit of 40 tons trucks over U.S. roads. Transportation of commercial SNF packages by truck requires a heavy haul truck (HHT) designation making long distance commercial SNF shipments utilizing trucks unfavorable. Barge transportation is another option that is available to transport commercial SNF, but the limited network of navigable waterways in the United States makes rail the preferred mode of transportation for SNF. However, it must be realized that not all reactor sites have rail lines, and even if rail lines are available, they might currently be inoperable. Therefore, for reactor sites with no operable rail lines, it is anticipated that barge or heavy haul trucks could be used for the initial transport leg until SNF shipments can be transferred to rail at a suitable transloading site. The general cycle of SNF transportation is anticipated to start from the railcar fleet maintenance facility (FMF). The rail cars depart for the pickup site from the FMF (with a transload leg if necessary). Once the SNF is picked up, the train travels to the drop site. Depending on if cask railcars, after dropping off their loaded casks, receive empty casks for maintenance, they are routed to a cask maintenance facility and subsequently back to the FMF. Figure 2 depicts the anticipated high-level process flow diagram for SNF shipments with a transload leg (if required).

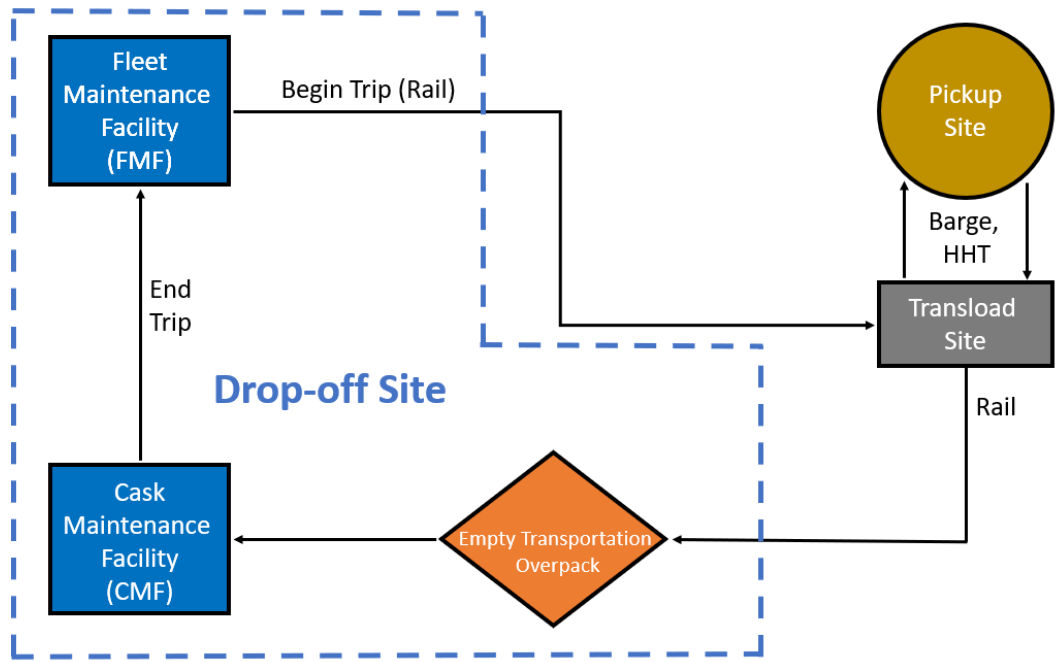

Figure. 2. A Potential Process Flow Diagram in System Simulations for loaded SNF Cask Railcars with a Transload Leg (If required) assuming the Drop Site, CMF, and FMF are collocated.

Systems analysis of back-end SNF logistics deals with modeling and changing the various input options available to the analyst such as shipping rates (number of packages), fuel pickup priority, transportation asset profiles (cask-carrying railcar, buffer railcar, escort railcar, or transportation overpacks), shipping dates, etc. Changing these variables enables the analyst to arrive at varying cost and time estimates to accomplish a SNF shipping campaign. Comparing the outputs of various scenarios provides the analyst the opportunity to evaluate the results of the choices made as part of the analysis.

**Next Generation System Analysis Model – Background**

The Next Generation System Analysis Model (NGSAM) is an agent-based discrete event simulation analytical tool [4]. The tool was developed from existing logistics models and utilizes the open-source agent-based modeling simulation platform Repast Simphony [5]. NGSAM was developed in response to key elements and recommendations of the Blue Ribbon Commission on America's Nuclear Future to help consider options for a future waste management system [6], including those which might include consolidated interim storage.

NGSAM's agent-based discrete event simulator allows the tool to model the back end of the fuel cycle down to the assembly level. Modeling the waste management system at this level of detail enables the analyst to perform detailed what-if analysis across the entire system. Another aspect of the tool is that most of the model logic is considered an input into the tool. The flexibility of having the model logic as input allows analysts to modify the logic without requiring a new version to be compiled and released. However, extremely complex logic, like SNF shipping allocation and canister packing, is coded directly into the tool code. The coding of this logic is to increase the performance and reduce the complexity of the logic required to perform the specific actions [7].

NGSAM has been developed to cover all possible pathways through the waste management system. Figure 3 shows the overview of various paths through the waste management system. To support the systems analysis, NGSAM allows analysts to answer questions based on parameters such as the SNF shipping allocation/acceptance logic; number and locations of consolidated interim storage facilities and monitored geologic repositories; repackaging; packaging; triple-purpose storage, transportation, aging, and disposal canisters; and non-canistered "bare fuel" transport in re-useable casks.

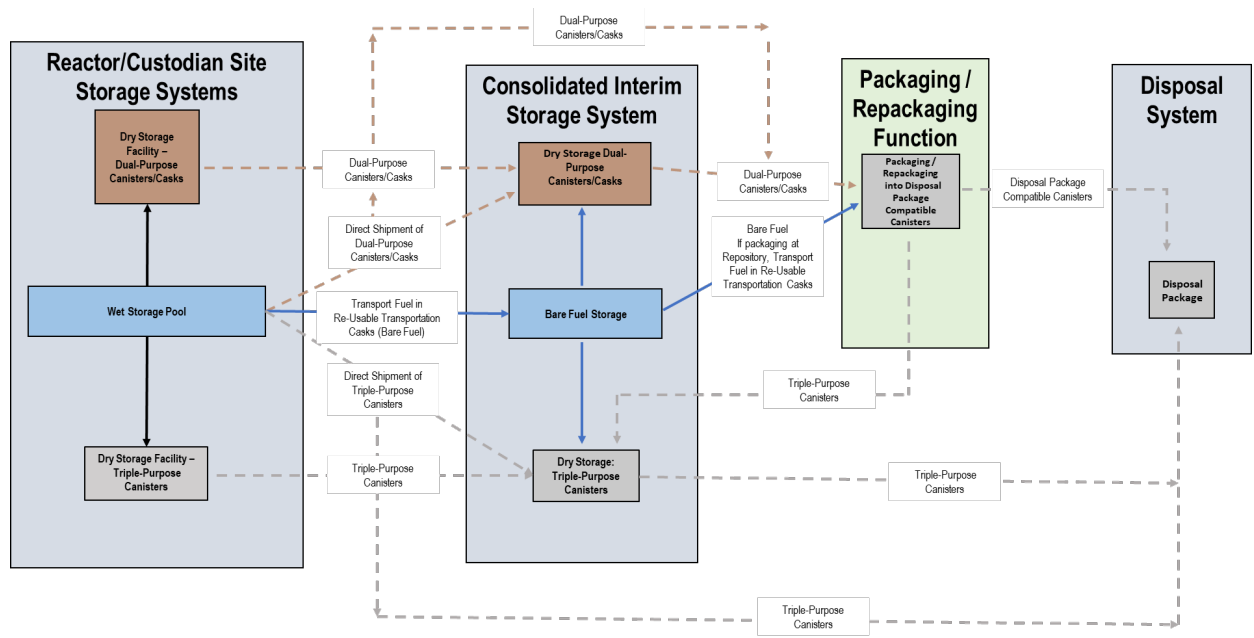

Figure. 3. Overview of the possible alternative SNF flowpaths which can be modeled in NGSAM.

Along with the input model logic, NGSAM utilizes data from multiple sources across the IWM program. The initial system state, including assembly and canister locations, is derived from the nuclear fuel data survey (Form GC-859) data collected from utilities by DOE. Additional data such as calculated heat profiles for individual assemblies calculated using the STANDARDS 5.0 spent fuel and analysis tool formerly known as the UNF-ST&DARDS are also used, along with projections of SNF from the STANDARDS Unified Database (UDB). NGSAM allows for multiple projections to be considered, addressing issues like early plant closure, and potential new reactor builds. The Performance Assessment of Strategic Options (PASO) tool is used to estimate timelines for the IWMS. NGSAM uses these timelines to inform when capabilities such as federal consolidated interim storage facilities (CISFs) could begin to operate during the simulation. Similarly, the Stakeholder Tool for Assessing Radioactive Transportation (START) is used to provide high-fidelity routes for shipments of SNF. NGSAM includes a default set of routes generated from the START tool from all SNF facilities to generic locations geographically distributed throughout the United States. NGSAM provides the analysts with the flexibility to override the default route sets with alternate routes generated to represent different possible scenarios.

Using the reference data stated above, along with the model logic, NGSAM simulates the waste management system down to the assembly level allowing each action in the system to be decided, performed, and recorded during the simulation. By executing the simulation at this level of fidelity, NGSAM provides users with detailed valuable data associated with analysis of various potential scenarios developed to examine the waste management system. By default, NGSAM generates a predefined set of comma-separated values (csv) files that analysts use to generate graphics and perform high-level statistical analysis. NGSAM also provides the utility of ad hoc report generation, which allows the user to indicate specific steps in the logic to record specific details. These ad hoc reports allow users to customize their output without requiring a separate compilation and release of the software. Finally, NGSAM records every action in the simulation in a file-based database. The database of results allows analysts to drill down into the simulation results to examine the specific behavior of individual elements within the system. For instance, NGSAM provides the ability to trace an assembly throughout its entire lifetime in the waste management system, from discharge to disposal, as depicted in Figure 4.

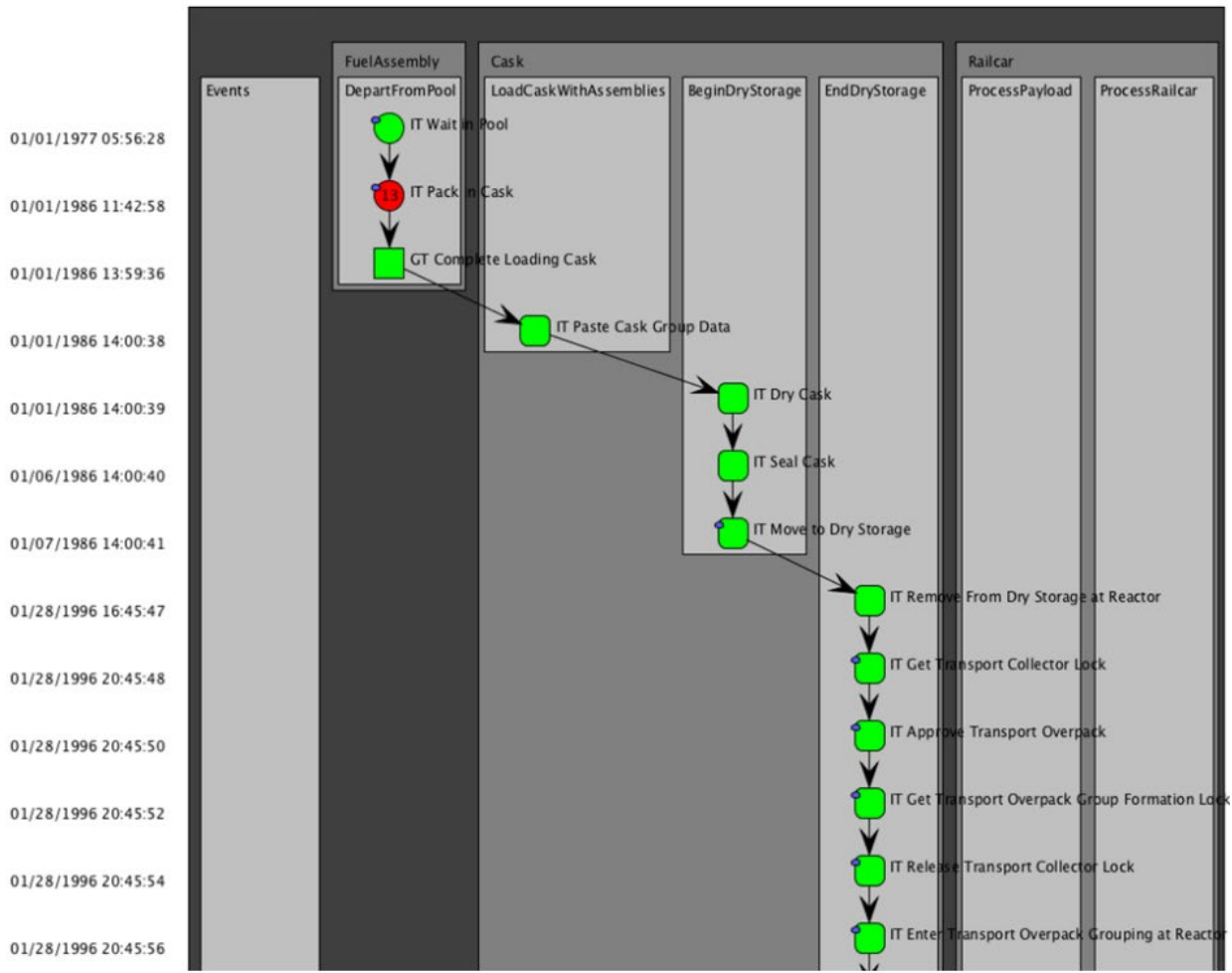

Figure. 4. A detailed report of the history of an assembly through a portion of the waste management system.

PNNL-SA-196362

NGSAM's ultimate goal is to provide system analysts with capabilities to model the IWMS and gain insights into SNF and HLW management alternatives including the impact of system choices, associated cost estimates, and development of integrated yet flexible approaches to inform decision makers.

**Stakeholder Tool for Assessing Radioactive Transportation – Background**

START is a web-based, decision-support tool developed by DOE to support the IWM program [8]. Its purpose is to provide the visualization and analysis of geospatial data relevant to planning and operating large-scale SNF and HLW transportation operations to consolidated interim storage and/or permanent disposal facilities.

At present, the primary transport method for these shipments is expected to be via rail, operating predominantly on mainline track. For many shipment sites, however, access to this network will typically require initial use of a local/regional (short line) railroad or involve transload operations where the access leg is a movement performed by heavy haul truck and/or barge. START can represent and analyze all of these transport options with each transportation network segment containing site-specific physical and operational attributes.

START users currently include Federal, State, Tribal, local government officials, and support contractors. DOE anticipates expanding access to nuclear utilities, transportation carriers, students, researchers, and other stakeholders. For this reason, START is designed to enable the user to develop a wide range of operating scenarios and performance objectives, with an emphasis on providing flexibility. In doing so, the tool makes extensive use of geographic information systems technology for performing spatial analysis and map creation.

In addition to the transportation network itself, START includes a variety of data layers that characterize populations, land use, and emergency response assets in proximity to potential shipment routes. Detailed attributes are provided for each point/link/polygon contained in each data layer that describe unique site-specific information. A suite of base map options is also provided for users to perform analyses and present results against the desired cartographic background.

When assessing routing options and risk attributes, users may select a shipment origin and destination from predefined locations provided via a drop-down menu or may identify any desired shipment location within the contiguous United States. When performing a routing analysis, the user may select from various routing criteria in finding a preferred solution. Once origin, destination, mode(s), and routing criteria are established, the user has the option to designate geographic locations to avoid as well as locations where the shipment is required to traverse through. This is often used to accommodate where there may be shipment size and/or weight limitations, or where route restrictions are in place [9]. Regarding the proximity of populations, land uses and emergency response assets to potential shipment routes, START considers two route buffer distances: 800 and 2,500 meters.

When a routing analysis is complete, START provides both summary (overall route) and detailed (route segment) performance measures and maps to support analysis and communication. Additionally, START allows users to report route analysis results according to political jurisdiction (e.g., by State, Tribal land, county, and congressional district).

PNNL-SA-196362

Route analysis results can be exported from START in a variety of arrangements, including as preformatted reports, shapefile, csv, and keyhole markup language (kml) file. This enables users to use START output in customized applications based on access to other in-house analysis and presentation tools. Users can also share routing analysis results with other users directly within the START platform.

Several other features are available in START to support user needs. Batch processing capability is available in situations where multiple routing analyses are being run concurrently (e.g., many shipment origins to a single shipment destination or many routing scenarios involving a specific shipment origin and destination). START can also accommodate photographic features, a convenient option when attempting to visualize potential obstacles along a potential route. Additionally, measurement tools (e.g., area, distance, elevation, and map coordinates) are available, as well as smart mapping capabilities from which spatial information can be filtered or presented by theme.

Beyond performing routing analyses, DOE can potentially utilize START for a number of other purposes, including (1) training preparations along routes (route proximity and response coverage provided by the fire, police, hospitals, and DOE Transportation Emergency Preparedness Program [TEPP] trained personnel), (2) stakeholder communication (visualization relative to locations of interest), (3) environmental analyses (transportation dose estimates), and (4) integration with other IWM software tools. It is worth mentioning that active verification and validation efforts are also underway to ensure critical START outputs are being independently verified such as the route length, buffer zone population (population density) as well as the incident free and public dose [10, 11].

**CROSS-COLLABORATIONS**

A CISF requires several tasks that need to take place before SNF can be accepted at the facility. Since the IWMS heavily relies on the cross-cutting interactions between areas such as transportation, storage, and disposal, efforts have been made to closely integrate these areas to develop a cohesive IWMS program. Several tasks were identified in fiscal year (FY) 2023 to understand the interdependencies of these various areas, some of which are detailed in this work.

**Travel Time and Sensitivities**

Travel time from an origin to a destination plays an important role in the turn-around time analysis of a transportation shipment. Currently, START and NGSAM both use fixed travel time; however, it is realized that in a real-world scenario, the travel time is expected to be a distribution with buffer times rather than a fixed value. These values are being actively investigated, and uncertainties are planned to be integrated in both the START reported values as well as in NGSAM. Contributions to the buffer time can also include routine activities such as transload site staging and preparation time as well as infrequent incidents, such as off-normal events and accidents that are bound to increase the travel time. Currently, data is reported only in the form of travel time when a route is run from an origin to a destination. However, there is potentially a need to report time both in terms of travel time as well as actual time since there might be some activities that do not require active travel but nevertheless accrue time, such as transload operations, inspection activities, crew change, etc.

**Development and Data Integration Initiatives**

START and NGSAM have two independent development teams that are responsible for these products. New versions of these tools are released periodically based on user inputs, verification and validation efforts, and sponsor needs. Since both these tools are anticipated to be used by decision makers to aid in the CISF design and development efforts, it was decided to allow the developers to gain access to the development environments for both these tools. Efforts are also being made to directly access the application programming interface (API) for the START tool rather than going through the START user interface to run routes. This could potentially allow users in NGSAM to directly run routes in START using the NGSAM user interface. As part of these efforts, custom, user-defined route import options were enabled in NGSAM as opposed to only predefined routes being used in prior releases of the tool. This is one of the first steps to seamlessly integrate the START generated outputs in NGSAM.

In addition, the START tool provides the user with extensive data on top of route length, travel time, and dose-related information. All this information currently generated by the START tool along a potential route is not stored or used for future analysis. However, it is anticipated that once a site for a federal CISF is identified, this information will be very useful in performing route readiness, environmental impact, and emergency response assessments and studies. In order to facilitate this, it is important to store all the START-generated data. Efforts are being made to understand the design efforts to enable the storage of all this data in the UDB. This is the same database that is used to store most of the data currently used by the NGSAM tool.

**CONCLUSIONS**

System analysts, NGSAM tool developers and START developers have been working closely for several years before formalizing this collaboration. START provides the SNF routing information for use in NGSAM. System analysts use the NGSAM tool to generate results (schedule, costs, infrastructure acquisitions, shipment rates, etc.). Subject-matter experts and system analysts work together to inform how NGSAM should model the waste management system. Specifically, this paper discussed the collaborations between the system analysis, NGSAM, and START teams and their accomplishments over the past year. Some examples of these efforts include travel time and sensitivities and development and data integration initiatives. Detailed examples of various tasks that have been performed by the team have been discussed. Continued efforts in this direction are expected in FY-24 and coming years.